\documentclass[final,5p,times,twocolumn]{elsarticle}

\usepackage{amsmath}
\usepackage{amssymb}
\usepackage{graphicx}
\usepackage{color}
\usepackage{hyperref}

\usepackage[mathscr,scaled=1.15]{urwchancal}
\DeclareFontFamily{OT1}{pzc}{}
\DeclareFontShape{OT1}{pzc}{m}{it}%
{<-> s * [1.15] pzcmi7t}{}
\DeclareMathAlphabet{\mathpzc}{OT1}{pzc}{m}{it}

\definecolor{purple}{rgb}{0.5,0,0.5}
\definecolor{blue}{rgb}{0.0,0,0.9}

\biboptions{sort&compress}

\journal{Physics Letters B}

\begin{document}

\begin{frontmatter}
\title{A Universal Constraint on the Infrared Behavior of the Ghost Propagator in QCD }

\author[PKU1,PKU2]{Fei Gao}
\author[PKU1,PKU2]{Can Tang}
\author[PKU1,PKU2,PKU3]{Yu-xin Liu}

\address[PKU1]{Department of Physics and State Key Laboratory of Nuclear Physics and Technology, Peking University, Beijing 100871, China}
\address[PKU2]{Collaborative Innovation Center of Quantum Matter, Beijing 100871, China}
\address[PKU3]{Center for High Energy Physics, Peking University, Beijing 100871, China}

\date{\today}

\begin{abstract}
With proposing a unified description of the fields variation at the level of generating functional,
we obtain a new identity for the quark-gluon interaction vertex based on gauge symmetry,
which is  similar to  the Slavnov-Taylor Identities(STIs) based on the Becchi--Rouet--Stora-Tyutin transformation.
With these identities, we find that in Landau gauge, the dressing function of the ghost propagator approaches to a constant as its momentum goes to zero, which provides a strong constraint on the infrared behaviour of ghost propagator.
\end{abstract}

\end{frontmatter}

\medskip

\noindent\textbf{{\emph{1. Introduction:}}}---Quantum Chromo-Dynamics(QCD) is a non-Abelian gauge theory,
of which the action  is built upon the SU(3) gauge symmetry.
However, the gauge symmetry of the action is broken after the gauge fixing in practical calculations.
Then, in a local covariant operator formalism introduced by Faddeev and Popov~\cite{Faddeev:1967fc},
the ghost and the anti-ghost fields are needed in the action.
Though the SU(N) gauge symmetry has been broken, a new symmetry called the Becchi--Rouet--Stora-Tyutin (BRST) symmetry is found in the complete action~\cite{Becchi:1974md}.
As a consequence, the Ward-Takahashi identities (WTIs)~\cite{Ward:1950xp, Takahashi:1957xn} based on the gauge symmetry are replaced by the Slavnov--Taylor identities (STIs)~\cite{Taylor:1971ff,Slavnov:1972fg}.
%
%

The BRST transformation  can be considered as  the generalization of the SU(N) gauge transformation by generalizing the infinitesimal quantity of the gauge transformation from real number field to the Grassmann number field.  For instance, the gauge transformation of the fermion field in the usual SU(N) gauge transformation reads $\delta\psi = -ig t^{a} \theta^{a} \psi$ with $\theta^{a}$ a real number, the respective BRST transformation is to replace the real number to Grassmann number, i.e. $\delta\psi = -ig t^{a} \lambda c^{a} \psi$ with $c^{a}$ the Grassmann number and $\lambda$ the phase angle. The field spanned by the Grassmann numbers is just the ghost field. The STIs can be obtained with the BRST transformation that keeps the action unchanged.

 Generally speaking, the variation of the fields does not need to be constrained by the symmetry of the action, instead, the variation can be employed at the level of generating function.  In the generating functional, the fields  are variables of the integral, and thus, the variation of the fields can be regarded as the replacement of the integral variables.
 Such a replacement can be taken to derive the STIs with the BRST symmetry of the action\cite{He:2009sj}.   As long as the functional integral converges, which is a natural demand for the generating functional,
 such kind technique can be employed.
 The relation derived from this variation might not be the charge conservation relation as $\partial_\mu J_\mu=0$, but with the additional source term, for instance, the mass term which breaks the chiral symmetry and leads to the relation: $\partial_{\mu}^{} J_{\mu5}^{} =m \Gamma_{5}^{}$.
 Here with such kind variation,  we derive a new WTI for the quark-gluon interaction vertex with the original gauge transformation.
 Meanwhile, the conventional STI from the BRST symmetry for the quark-gluon interaction vertex still holds.
 With these two identities, the ghost propagator can be strongly constrained.

People are interested in the infrared behaviour of the ghost propagator for a long time.
Kugo and Ojima have firstly proposed an infrared-enhanced ghost propagator
whose dressing function is divergent~\cite{Kugo:1979gm,Hata:1981nd}.
However, the assumption of the asymptotic fields has been employed in their derivations which might not be appropriate in QCD.
Gribov and Zwanziger have also conjectured an infrared enhanced ghost propagator by considering the overcompensation of the configurations close to the Gribov horizon~\cite{Gribov:1977wm,Zwanziger:1989mf}.
Nevertheless, a refinement of the Gribov-Zwanziger action has recently been proposed~\cite{Dudal:2012PRD,Zwanziger:2013PRD}. This improvement includes the condensates as the dynamical effects, and obtains that the ghost propagator is infrared-finite.
Besides, the numerical results of the recent calculations including those in the Dyson-Schwinger equations of QCD~\cite{Aguilar:2012PRD,Ayala:2012PRD,Strauss:2012PRL} and the lattice QCD simulations~\cite{Bogolubsky:2009PLB,Boucaud:2010PRD,Bowman:2004PRD,Cucchieri:2012PRD,Oliveira:2011JPG} seem to be more supportive to an infrared-finite ghost propagator.
Though the numerical computations are convincible, a model-independent analytic proof is still needed.

In this article, we derive the WTI for the quark-gluon interaction vertex which is parallel to the STI.
Combining the WTI and the STI for the quark-gluon interaction vertex together,  we give a  universal conclusion on the infrared behaviour of the ghost propagator, that is, in Landau gauge, the dressing function of the ghost propagator at zero momentum becomes constant.
The infrared behaviour of the ghost propagator is then clarified.

\medskip

\noindent\textbf{{\emph{2. Derivations and Discussions:}}}---In the local covariant operator formalism, the action of QCD is:
\begin{equation}
S = \int d^4x \mathcal{L}_{QCD}^{} \, ,
\end{equation}
with

\begin{eqnarray}
\begin{split}
 \mathcal{L}_{QCD}^{} = \; &  \bar{\psi}(- \partial\!\!\!/ + m) \psi
 - ig\bar{\psi} \gamma_{\mu}^{} t^{a} \psi A^{a}_{\mu}     \\
& + \frac{1}{2}A^{a}_{\mu} \Big{(} -\partial^{2} \delta_{\mu\nu}^{} - \big{(} \frac{1}{\xi} - 1 \big{)}
       \partial_{\mu}^{} \partial_{\nu}^{} \Big{)} A^{a}_{\nu}    \quad  \\
& - gf^{abc} (\partial_{\mu} A^{a}_{\nu}) A^{b}_{\mu} A^{c}_{\nu}  \\
%
& +\frac{1}{4} g^{2} f^{abe} f^{cde} A^{a}_{\mu} A^{b}_{\nu} A^{c}_{\mu} A^{d}_{\nu}      \\
& + \bar{c}^{a} \partial^{2} c^{a}
    + gf^{abc}\bar{c}^{a} \partial_{\mu}^{} (A^{c}_{\mu} c^{b} )\, ,
\end{split}
\end{eqnarray}
where $\bar{\psi}$, $\psi$, $A^{a}_{\mu}$, $\bar{c}$, $c$ is the anti-quark, quark, gluon, anti-ghost and ghost field, respectively. $t^{a}$ and $f^{abc}$ are the generators and the structure constants of the SU(3) gauge group, respectively.
This action is no longer gauge invariant because the gauge fixing term changes corresponding to the gauge transformation. The action is only invariant under the BRST transformation defined by:
\begin{eqnarray}
\begin{split}
\delta \Psi=-igt^{a} c^{a} \psi\lambda \, , \qquad
&  \delta A^{a}_{\mu} = (\delta^{ab} \partial_{\mu}^{} + gf^{abc} A^{c}_{\mu}) c^{b} \lambda\, , \quad \\
\delta c^{a} = - \frac{g}{2}f^{abc} c^{b} c^{c} \lambda \, ,           \quad
& \delta\bar{c}^{a} =\frac{1}{\xi}\partial_{\mu}^{} A^{a}_{\mu}\lambda \, .
\end{split}
\end{eqnarray}
From the BRST symmetry, the STI for the quark-gluon interaction vertex can be written as~\cite{Eichten:1974et,Alkofer:2008tt}:
\begin{eqnarray}
\label{eq:sti}
\begin{split}
G^{-1}(k^2)k_\mu \Gamma^a_\mu(p,q) = & \; \big{[} g t^{a} - B^{a} (k,q) \big{]} S^{-1}(p) \\
& -S^{-1}(q) \big{[} g t^{a} - B^{a} (k,q) \big{]}  \, , \quad
\end{split}
\end{eqnarray}
where $G(k^{2})$ is the dressing function of the ghost propagator,
$D^{ab}(k)=-\delta^{ab}G(k^{2})/k^{2}$, and $B^{a} (k,q)$ the ghost-quark scattering kernel.
This STI contains the ghost propagator and the ghost-quark scattering kernel, which is unknown.
People usually employ the so called Abelian approximation that eliminates the ghost effect during computations.
In Abelian approximation, the STI degenerates into the WTI:
$$k_{\mu}^{} \Gamma_{\mu}^{}(p,q) = g S^{-1}(p) - S^{-1}(q)g \, . $$
With this constraint, people usually take the Ball-Chiu vertex~\cite{Ball:1980ay}
as the longitudinal part of the quark-gluon interaction vertex~\cite{Roberts:1994dr,Qin:2013mta,Gao:2016qkh}.

As we analyzed above,  this identity is based on the BRST transformation to guarantee the action invariant. However, the transformation does not need to coincide with this constraint beforehand.
To understand this, we resort to the variation of the generating functional.
In the path integral scheme, the generating functional is  defined as:
\begin{equation}
Z[J] = \int \mathcal{D} \phi_{i}^{} \exp\{-\int d^{4}x(\mathcal{L}_{QCD}^{} +J_{i}^{} \phi_{i}^{} ) \} \, ,
\end{equation}
with  $\phi$ standing for any field in the Lagrangian of QCD.
If the generating functional converges, the functional is unchanged under the replacement of the integral variables, that is:
\begin{eqnarray}
\begin{split}
&\int \mathcal{D}\phi^{\prime} \exp\Big{\{} -\int d^{4} x (\mathcal{L^{\prime}}_{QCD}^{} + J_{i}^{} \phi^{\prime}_{i} ) \Big{\}}           \\
= \;  & \int \mathcal{D} \phi \exp \Big{\{} -\int d^{4} x (\mathcal{L}_{QCD}^{} + J_{i}^{} \phi_{i}^{} ) \Big{\}} \, .
\end{split}
\end{eqnarray}

The replacement is naive, but will lead to nontrivial results.
At first, we can implement it to derive the axial-vector Ward-Takahashi identities (AV-WTIs).
Considering the replacement:
\begin{eqnarray}
\begin{split}
\psi^{\prime}  = & \psi - ig \frac{\tau^{i}}{2} \gamma_{5}^{} \theta^{i} \psi \, , \\
\bar{\psi^{\prime}}  = & \bar{\psi} - ig \bar{\psi}\frac{\tau^{i}}{2}\gamma_{5}^{} \theta^{i} \, ,
\end{split}
\end{eqnarray}
where $\tau^i/2$ is the generator of the group demonstrating the flavour symmetry.
Under this transformation, only the terms $\bar{\psi}(-\partial\!\!\!/+m)\psi$  and  the external sources change in the generating functional. With the above mentioned principle that the functional is invariant under the replacement of the integral variables, we can get straightforwardly
\begin{eqnarray} \label{eq:wt1}
\begin{split}
& \Big{\langle} \theta^{i} \partial_{\mu}^{} (\bar{\psi}(x) \gamma_{\mu}^{}  \frac{\tau^{i}}{2} \gamma_{5}^{} \psi(x)) \Big{\rangle} - 2m \Big{\langle} \theta^{i} \bar{\psi}(x) \frac{\tau^{i}}{2} \gamma_{5}^{} \psi(x)  \\
& \; \; + \bar{\psi}(x) \frac{\tau^{i}}{2} \gamma_{5}^{} \theta^{i} \eta(x)
       + \bar{\eta}(x) \frac{\tau^{i}}{2}\gamma_{5}^{} \theta^{i} \psi(x) \Big{\rangle} =0 \, ,
\end{split}
\end{eqnarray}
with $\eta$, $\eta^{\prime}$ being the external sources of $\bar{\psi}$ and $\psi$, respectively.
Deviating it by $\frac{\delta^{2}}{\delta\eta(x_{1}^{})\delta\bar{\eta}(x_{2}^{})}$ and converting into the momentum space, Eq.~(\ref{eq:wt1}) reads
\begin{eqnarray} \label{eq:avwt}
\begin{split}
- i k_{\mu}^{} \Gamma^{i}_{5\mu}(p,q) = & S^{-1}(p) \gamma_{5}^{} \frac{\tau^{i}}{2}
+ \gamma_{5}^{} \frac{\tau^{i}}{2}S^{-1}(q) \\
&  - 2m \Gamma^{i}_{5} (p,q) \, ,
\end{split}
\end{eqnarray}
where $k=p-q$ is the momentum transfer, $S^{-1}$ is the inverse of the quark propagator,  $\Gamma^{i}_{5\mu}$ and $\Gamma^{i}_{5}$ stand for the isovector axial-vector vertex, the isovector pseudoscalar vertex, respectively.
We can see that it is exactly the AV-WTI with a mass term which is the source term breaking the symmetry.
The respective vector Ward-Takahashi identity (V-WTI) can be obtained via the similar transformation
\begin{eqnarray}
\begin{split}
\psi^{\prime}  = & \psi - ig \frac{\tau^{i}}{2} \theta^{i} \psi   \, , \\
\bar{\psi^\prime} = & \bar{\psi} + ig \bar{\psi} \frac{\tau^{i}}{2} \theta^{i} \, .
\end{split}
\end{eqnarray}
It is easy to find that only the terms $-\bar{\psi}\partial\!\!\!/\psi$  and  the external sources change in the generating functional. With the similar procedure, we get
\begin{eqnarray} \label{eq:wt2}
\begin{split}
& \Big{\langle} \theta^{i} \partial_{\mu}^{} (\bar{\psi}(x) \gamma_{\mu}^{} \frac{\tau^{i}}{2} \psi(x)) \Big{\rangle} \\
+ & \; \Big{\langle} \bar{\psi}(x) \frac{\tau^{i}}{2} \theta^{i} \eta(x)
- \bar{\eta}(x) \frac{\tau^{i}}{2} \theta^{i} \psi(x) \Big{\rangle}  = 0 \, .
\end{split}
\end{eqnarray}
Deviating it and converting the result into momentum space, we have
\begin{equation} \label{eq:vwt}
- i k_{\mu}^{} \Gamma^{i}_{\mu} (p,q) = S^{-1}(p) \frac{\tau^{i}}{2} - \frac{\tau^{i}}{2} S^{-1}(q) \, ,
\end{equation}
where $\Gamma^{i}_{\mu}$ is the isovector vector vertex.
These two vertices are both color-singlet vertices.
The axial-vector vertex contains  pion.
Therefore, the AV-WTI has been widely used in QCD,
succeeded to predict pion as the Goldstone boson and also the particle that reveals the dynamical chiral symmetry breaking~\cite{Maris:1997hd}.
The V-WTI has also been employed as the conditions that must be satisfied when computing the meson's form factor~\cite{Maris:1999bh}.
It can be seen that the  variation replacement of the generating functional can be taken to derive the identities no matter the symmetry has been broken or not.

Now considering the color gauge transformation for $\psi$ and $\bar{\psi}$ which reads
\begin{eqnarray}
\begin{split}
\psi^{\prime} = \; & \psi - ig t^{a} \theta^{a} \psi    \, , \\
\bar{\psi^{\prime}} = \; & \bar{\psi} + ig \bar{\psi} t^{a} \theta^{a} \, ,
\end{split}
\end{eqnarray}
which could also be regarded as the variables' replacement in the generating functional integral.
The Jacobian determinant of this transformation is
\begin{eqnarray}
\begin{split}
J_{ij}^{} \bar{J}_{mn}^{} = \; & \det(\delta_{ij}^{} - ig t^{a}_{ij} \theta^{a})
\det( \delta_{mn}^{} + ig t^{a}_{mn} \theta^{a})  \qquad \\
 = \; & 1 + \mathcal{O}(\theta^2) \, .
\end{split}
\end{eqnarray}
Under this transformation, only the terms $\bar{\psi}\partial\!\!\!/\psi$,
$g\bar{\psi}\gamma_{\mu}^{} t^{a} \psi A^{a}_{\mu}$ and the external sources change in the generating functional, we get directly then
\begin{eqnarray}
\begin{split}
 \big{\langle} \partial_{\mu}^{} (\bar{\psi}(x) \gamma_{\mu}^{} t^{a} \psi(x) )
+ g f^{abc} A^{b}_{\mu} \bar{\psi}(x) \gamma_{\mu}^{} t^{c} \psi(x)  & \\
 + \bar{\psi}(x) t^{a} \eta(x)  -  \bar{\eta}(x) t^{a} \psi(x) \big{\rangle} & = 0 \, . \qquad
\end{split}
\end{eqnarray}
Deviated by $\frac{\delta^2}{\delta\eta(x_1)\delta\bar{\eta}(x_2)}$, the expression becomes:
\begin{eqnarray} \label{eq:wt3}
\begin{split}
& \big{\langle} \partial_{\mu}^{} (\bar{\psi}(x) \gamma_{\mu}^{} t^{a} \psi(x)) \psi(x_{2}^{}) \bar{\psi}(x_{1}^{}) \big{\rangle} \\
& \;\; + \big{\langle} g f^{abc} A^{b}_{\mu} \bar{\psi}(x) \gamma_{\mu}^{} t^{c} \psi(x) \psi(x_{2}^{}) \bar{\psi}(x_{1}^{} ) \big{\rangle} \\
 = \; & i\delta^{4}(x - x_{2}^{}) \big{\langle} \psi(x) t^{a} \bar{\psi} (x_{1}^{}) \big{\rangle}  \\
       & \; \; - i\big{\langle} \psi(x_{2}^{}) t^{a} \bar{\psi}(x) \big{\rangle} \delta^{4} ( x - x_{1}^{} ) \, .
\end{split}
\end{eqnarray}
This identity is exact without including any approximation.
The second term in Eq.~(\ref{eq:wt3}) is unusual,
since it owns very similar structure with that in the quark propagator's self-energy.
Expanding this term in the momentum space we can obtain
\begin{eqnarray}
 & & \int d^{4} k e^{-ikx} \big{\langle} g f^{abc} A^{b}_{\mu} \bar{\psi}(x) \gamma_{\mu}^{} t^{c} \psi(x) \psi(p) \bar{\psi}(q) \big{\rangle} \nonumber \\
 & =\; & \delta^4(k-p+q) \frac{1}{2} f^{abc}f^{bcd} \nonumber \\
 & & \times S(p) \Big{\{} \Sigma^{d} (p) - \Sigma^{d} (q)
    + \int d^{4} k^{\prime} D_{\mu\nu}^{} \gamma_{\mu}^{} H^{d}_{6} (k,p,q) \Gamma_{\nu}^{} \notag \\
 & & \;\; + \! \int \! d^{4} k^{\prime}  \Sigma^{d}(\! p-k^{\prime} \!) S(\! p-k^{\prime} \!)
 K(\! p-k^{\prime} \! ,\! q- k^{\prime} \!) S(\! q-k^{\prime} \! )  \notag  \\
 & & \;\; - \! \int \! d^{4} k^{\prime} \Sigma^{d} (\! q-k^{\prime} \!) S(\! q-k^{\prime} \!)
 K(\! q-k^{\prime} \! , \! p-k^{\prime} \!) \notag   \\
 & & \qquad \qquad S(\! p-k^{\prime} \! ) \Big{\}} S(q)   \, ,
\end{eqnarray}
where
$$\displaystyle \Sigma^{d}(p) = g^{2} \int d^{4} k^{\prime} D_{\mu\nu}^{}(k^{\prime}) \gamma_{\nu}^{} S(p-k^{\prime} ) \Gamma^{d}_{\mu}(p, p-k^{\prime}) $$
is the self energy of the quark  without the sum of color indices,
$K(p, q)$ is the four-quark scattering kernel,
$H^{d}_{6}(k, p, q)$ is in terms of the six-quark Green function.
Converting the relation into the momentum space, we have
\begin{eqnarray}
\begin{split}
k_{\mu}^{} \tilde{\Gamma}^{a}_{\mu} (p, q) = & \;\, g t^{a}S^{-1}(p)\! - \! S^{-1}(q) g t^{a} \! \\
&- \! \frac{N^2_c}{(N_{c}^{2} \! - \! 1)} \Big{(} g t^{a} \Sigma^{}(p)-\Sigma^{}(q) g t^{a}  \!\qquad \\
& + t^{a} \overline{K}(p,q)
    + \overline{K}(q,p) t^{a}  + \overline{H}^{a}_{6} (p,q) \Big{)} \, ,
\end{split}
\end{eqnarray}
where
$$ \overline{K}(p,q) = \int d^{4}k^{\prime} \Sigma(p-k^{\prime}) S(q-k^{\prime} ) K(p-k^{\prime} , q-k^{\prime} ) S(q-k^{\prime} ) \, , $$
$$\overline{H}^{a}_{6} (p, q) = \int d^{4} k^{\prime} D_{\mu\nu}^{} \gamma_{\mu}^{}  H^{a}_{6}(k,p,q) \Gamma_{\nu}^{} \, . $$
The definition of the quark-gluon interaction vertex $\tilde{\Gamma}^{a}_{\mu}$  is a little different from above, which is only the vertex  that the gluon couples directly with quarks,
the difference from the complete vertex is a six-point scattering kernel $B_{6}^{}(p,q)$ in
$ \big{\langle} \bar{\psi}(x) \gamma_{\mu}^{} t^{a} \psi(x) \bar{c}^{e} (x) c^{f} (0) \psi(x_{2}^{}) \bar{\psi}(x_{1}^{}) \big{\rangle} $ with four quarks and two ghosts as shown in Ref.~\cite{He:2009sj}.
With this correction we get:
\begin{eqnarray}
\begin{split}
 &\frac{1}{ 1 - B_{6}^{} (p,q) } k_{\mu}^{} \Gamma^{a}_{\mu}(p,q)  \\
\!\!\! = &  \, g t^{a}S^{-1}(p)\! - \! S^{-1}(q) g t^{a} \! - \! \frac{N^2_c}{(N_{c}^{2} \! - \! 1)} \Big{(} g t^{a} \Sigma^{}(p) \! - \! \Sigma^{}(q) g t^{a}  \\
& + g t^{a} \overline{K}(p,q) + \overline{K}(q,p) g t^{a} + \overline{H}^{a}_{6}(p,q) \Big{)} \, .
\end{split}
\end{eqnarray}
%

This is  another identity for the quark-gluon interaction vertex,
which represents the relation between the gauge current and the symmetry breaking source,
just like the case of AV-WTI.
Noticing that it is exact in QCD, and the STI, Eq. (\ref{eq:sti}), still holds at the same time.
These two sets of identities stand for two different transformations,
one is the gauge transformation and the other is the BRST transformation. The two transformations are respective to different phase spaces, and then the two sets of identities together describe the complete structure  in QCD at the level of generating functional.

Moreover, if combining the WTI for the quark-gluon interaction vertex we derived here with Eq. (\ref{eq:sti}), it gives a strong constraint for the ghost propagator.
In Landau  gauge, it has been shown~\cite{Taylor:1971ff} perturbatively that $B^{a}(k,q)\rightarrow 0$
for $k \rightarrow 0$, but the nonperturbative effect might make it deviate against 0. The analysis in Ref.~\cite{Alkofer:2008tt} has shown that the  function
$$t^aH(p,q)=gt^a-B^a(p,q)$$
should be infrared finite, and thus it may change the value of $G(k^2=0)$  quantitatively, but not qualitatively. Besides,
the six-point kernels $B_{6}^{}$ and $\bar{H}^{a}_{6}$ are the higher order corrections, which will arise some more dressing effect on the quark-gluon interaction vertex but will not change the behaviour of the ghost propagator quantitatively.
In practical  computations (e.g., Refs.~\cite{Qin:2013mta,Fischer:2008uz}),
such higher order corrections are usually neglected.
After this, if we take $p$ and $q$ both tend to be zero,  we get the value of the dressing function of the ghost propagator at zero momentum in Landau gauge as
\begin{equation}
G(k^{2} =0) =  \frac{1/H(0,0)}{1+\big{(} N_{c}^{2}/(N_{c}^{2}-1) \big{)}
\Gamma(0) } \, .
\end{equation}
with
$$ \Gamma(p-q) = \frac{tr[k\!\!\!/\big(\Delta \Sigma(0)+\int d^{4} k^{\prime}  \big{(} \Delta \Sigma(k)) t^{a} \big{)} SKS\big)]}{tr [k\!\!\!/k_{\mu}^{} \Gamma_{\mu}^{a}]},$$
where only the Lorentz indices are  taken  into account in the trace and $\Delta \Sigma(k)= \Sigma^{}(p-k^{\prime} ) - \Sigma^{}(q-k^{\prime})$. If only considering the leading order of $k\!\!\!/$ in the above formula, we can see that the leading order of  $ \Sigma^{}(p-k^{\prime} ) - \Sigma^{}(q-k^{\prime} )$ is proportional to $(A(k^{\prime2})-1)k\!\!\!/$ with $A(k^{\prime2})$ being the vector part of the quark propagator. The remanent part  $t^a k\!\!\!/+t^a k_\mu\int d^{4} k^{\prime}  \gamma_\mu SKS$  is just the  expansion for the quark-gluon interaction vertex in the Dyson-Schwinger equations and can be equivalently rewritten as $k_{\mu}^{} \Gamma_{\mu}^{a}$.
Taking $A(0)$ as the approximation of $A(k^{\prime2})$, $\Gamma(0)$ in the above formula can be simplified as $A(0)-1$. With typical numerical value obtained previously,  $A(0)=1.7$, we can get the value of the ghost dressing function at zero momentum under the SU(3) gauge symmetry as $G(0)=0.56/H(0,0)$.
This result  demonstrates the infrared behavior of the ghost propagator.
It proves that the ghost propagator is in the so called decoupling solution whose dressing function  becomes constant at the deep infrared limit.
It is known that  there may exist two solutions for the ghost and gluon propagators depending on the boundary condition~\cite{Fischer:2008uz}, and there can be another solution of which the dressing function is infrared-enhanced if taking the value of $A(0)$ as a  small value down to $1/9$.
However, this has not yet been obtained in the numerical calculation on the quark propagator with realistic parameters.
Therefore,  such a relation reveals that the quark-gluon interaction part serves as the boundary condition
and  gives strong constraints on the properties of the gauge fields.

\medskip

\noindent\textbf{{\emph{3. Summary:}}}---In our sense, the relations of the Green functions have broader forms which need not to be limited by the symmetry of the original action. For instance, even though the isovector axial-vector  symmetry can be broken by the mass term in the action, the AV-WTI with source term can still be obtained.
Similarly, even though the gauge symmetry has been broken by the gauge fixing term in QCD's local covariant action, the WTIs based on the gauge symmetry can still be obtained with some additional source terms, which is another set of identities parallel to the STIs based on the BRST symmetry just like the relation between the AV-WTIs and the V-WTIs.

In this paper we give a unified description for the symmetry of the Green functions, i.e. at the level of the generating functional. The key point is that the field variation in the generating functional is regarded as the replacement of the integral variables.
Therefore, the replacement can be arbitrary without being constrained by the symmetry.
Such replacement under the gauge transformation can be easily implemented in practical calculations.
We get, in turn, another WTI with source term for the quark-gluon interaction vertex.
By combining the presently obtained WTI  with the STI,
we find a relation that strongly constrains the infrared behavior of the dressing function of the ghost propagator.
The relation proves that in Landau gauge, the dressing function of the ghost propagator becomes constant at zero momentum.

\medskip

We thank Prof. Si-xue Qin, Prof. Lei Chang and Prof. Craig D. Roberts for useful discussion.
The work was supported by the National Natural Science Foundation of China under Contract No. 11435001; the National Key Basic Research Program of China under Contracts No.~G2013CB834400
and No.~2015CB856900.

\end{document}